\documentclass[aps,prb,twocolumn,amsmath,amssymb,superscriptaddress]{revtex4}

\usepackage[T1]{fontenc}
\usepackage[latin1]{inputenc}
\usepackage{graphics}
\usepackage{amssymb}
\usepackage{amsmath}
\usepackage{overpic}

\begin{document}

\title{Hopping-resolved electron-phonon coupling in bilayer graphene}

\author{E. Cappelluti}

\affiliation
{Instituto de Ciencia de Materiales de Madrid,
ICMM-CSIC, Cantoblanco, E-28049 Madrid, Spain}

\affiliation
{Institute for Complex Systems (ISC), CNR, U.O.S. Sapienza,
v. dei Taurini 19, 00185 Roma, Italy}

\author{G. Profeta} 
\affiliation { Dipartimento di Fisica, Universit\`a dell'Aquila, Via Vetoio 10, I-67100 L'Aquila, Italy}
\affiliation{ CNR-SPIN Via Vetoio 10, I-67100 L'Aquila, Italy}

\date{\today}

\begin{abstract}
In this paper we investigate the electron-phonon coupling
in bilayer graphene, as a paradigmatic case for multilayer graphenes
where interlayer hoppings are relevant.
Using a frozen-phonon approach within the context
of Density Functional Theory (DFT) and using different optical phonon
displacements we are able to evaluate quantitatively
the electron-phonon coupling $\alpha_i$
associated with each hopping term $\gamma_i$.
This analysis also reveals a simple scaling law between
the hopping terms $\gamma_i$ and the electron-phonon coupling
$\alpha_i$ which goes beyond the specific DFT technique employed.

\end{abstract}


\maketitle

\section{Introduction}

Since its discovery,
a formidable amount of work has been devoted to investigate
the electronic and structural properties of
single-layer and multi-layer graphenes.
The electron-phonon interaction has in particular attracted
a huge interest for its role in controlling the charge
transport,\cite{chen,fratini,hwang,efetov,ochoa}
for providing a powerful interface between electronic
and structural properties,\cite{voz,lopes,dejuan,mucha}
and also because phonon resonances
in Raman and infrared spectroscopies,
triggered by the electron-phonon interaction,
represent a useful way to characterize graphenic samples
and to reveal interesting unconventional
effects.\cite{ferrari,pisana,yan1,yan2,malard,kuzmenkoFano,tang,lilui}

\begin{figure}[b]
\includegraphics[scale=0.4,clip=]{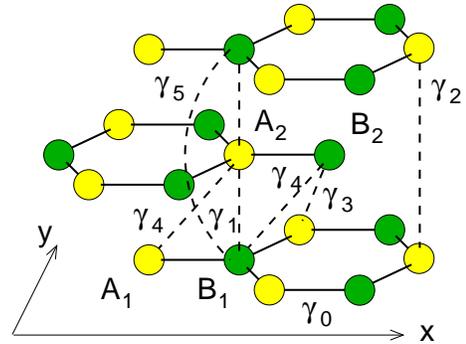}
\caption{(color online) (a) Atomic structure of multilayer graphene
with Bernal stacking showing the relevant hopping terms $\gamma_i$.
Atoms B1 and A2, connected by vertical $\gamma_1$,
denoted by darker colors, contain also a local
crystal field potential.}
\label{f-sketch}
\end{figure}

On the theoretical level, tight-binding (TB) models
are of fundamental importance since they
have been shown to catch almost all the electronic features
in this systems.
The vast majority of works based on TB in literature\cite{review}
employ a simple two-parameter TB model, where
only the nearest neighbor in-plane hopping $\gamma_0$
and the nearest neighbor vertical hopping $\gamma_1$
are considered, although, when needed, higher order
TB terms are included to reproduce more detailed
features.\cite{zhang,koshino,kuzmenko2L,li,avetisyan,mcd,zou,lui,apalkov}
Most important in bilayer graphene,
borrowing the terminology from bulk graphite,\cite{dresselhaus}
on the atoms B1 and A2, and the hopping terms
$\gamma_3$, operative between the atoms
A1-B2, and $\gamma_4$ between atom couples A1-A2 and B1-B2
(see Fig. \ref{f-sketch}). Such terms represent thus the basilar
ingredients to build a TB model in multilayer graphene with
both Bernal (ABAB\ldots) and rhombohedral (ABCA\ldots) stacking.
For instance, the hopping term $\gamma_3$ was shown to be related
to the trigonal warping and,\cite{dresselhaus} in bilayer systems, to the generation
of new Dirac points at finite momentum close to the K point.

Tight-binding models are also widely employed to investigate
the electron-phonon interaction. Focusing on the
single-layer graphene, the most relevant in-plane lattice vibrations
are related to the modulation of the nearest neighbor hopping
$\gamma_0$.\cite{review} Within this context, for instance, the optical properties
on the $E_{2g}$ phonon band at $\omega \approx 0.2$ eV
have been throughout investigated\cite{ando1,ando2,ando3,gava,our}
as well as the effects
on the electronic structure of the
long wavelength acoustic modes associated with the
ripples.\cite{ochoa,voz,dejuan}
Modeling of the electron-phonon interaction in multilayer graphene
is also commonly discussed on the basis of the modulation
of the $\gamma_0$ hopping. Among other things,
this kind of analysis was useful to show the robustness of
the Dirac points\cite{notedirac} upon lattice distortions,
in single- as well in multi-layer systems.\cite{manes,samsonidze}

Alternative to tight-binding model, Density Functional Theory (DFT)
calculations permit to include all the different kinetic
(e.g. hopping) terms at the same level.
It also permits to provide a {\em quantitative} estimate
of the electron-phonon coupling.
Pivot in this context is the concept of {\em deformation
potential}, {\em i.e.} the shift of the electronic levels upon
a frozen phonon lattice distortion, which is strictly
related to the magnitude of the electron-phonon interaction.\cite{khan}
In the context of graphenic materials, frozen phonon DFT
calculations were employed to quantify in single layer
systems the electron-phonon coupling associated to the
modulation of the $\gamma_0$ term upon a lattice
distortion $u$.\cite{lazzeri,piscanec}
It can be shown indeed that the in-plane optical mode $E_{2g}$
induces a linear splitting $\Delta \epsilon$ of the Dirac states at K
point, $\Delta \epsilon \propto 6 \alpha u$, where $\alpha$
is related to the linear coupling of the electronic states with the
${\bf q}=0$ $E_{2g}$ mode.\cite{lazzeri,samsonidze}
In a TB model, defining $u$
as {\em displacement per atom}, one gets $\alpha=d\gamma_0/du$.
DFT calculations in single-layer graphene obtain
$\alpha=4.5$ eV/\AA.\cite{lazzeri}
A similar value $\alpha \approx 4.4$ eV/\AA\, is found also
in graphite\cite{lazzeri,piscanec} where the linear energy splitting
$\Delta \epsilon$
at the H point (where the interlayer hopping is unaffective)
can be shown to be uniquely related
to $\alpha=d\gamma_0/du$.
In both cases, in graphene and graphite, a GW theory leads
to slight larger values $\alpha=5.1-5.3$ eV/\AA.\cite{lazzeri}

Despite large effort has been devoted thus in literature
to study the electron-phonon interaction related to
the $\gamma_0$ term, virtually no work has been addressed so far
to provide a quantitative estimate of the electron-phonon
coupling associated with the modulation of the other hopping terms.
A quantitative insight on this issue, on the other hand,
becomes increasing important because of the role of such terms
to many effects, from the establishment of unconventional anisotropic
phases in strained bilayer systems\cite{mucha,mariani}
to the evaluation of the optical properties of the
in-plane and out-of-plane phonon mode
in multilayer systems and in graphite.\cite{manzardo}

Aim of the present paper is to fill this gap and to provide,
with a first-principle DFT calculation,
a quantitative study of the electron-phonon coupling
associated with the modulation of other main hopping terms,
both for in-plane and for the out-of-plane vibrations.
We address this issue focusing on the optical phonon modes
at ${\bf q}=0$ in bilayer graphene.
The modulation of each hopping term with the relative distance,
however, permits to provide a generalization of the present results
at any finite ${\bf q}$.

\section{Frozen phonon analysis}

In this paper we consider single-layer and bilayer graphene
with typical Bernal stacking.
We take the in-plane nearest-neighbor carbon-carbon distance
$b=1.42$ \AA\,
($a=2.46$ \AA\, the lattice constant),
and the interlayer distance $d=3.35$ \AA.
Such lengths rule  thus the magnitude
of the in-plane hopping term $\gamma_0$
and out-of-plane hopping terms $\gamma_i$ on the relative distance
of the corresponding atoms.
For sake of simplicity, we denote with $b_i$ the distance
associated with each hopping term in the undistorted structure,
namely $b_0=|{\bf R}_{\rm A1}-{\bf R}_{\rm B1}|$,
$b_1=|{\bf R}_{\rm B1}-{\bf R}_{\rm A2}|$,
$b_3=|{\bf R}_{\rm A1}-{\bf R}_{\rm B2}|$,
$b_4=|{\bf R}_{\rm A1}-{\bf R}_{\rm A2}|$.
We assume that on the local scale
the hopping terms $\gamma_i$ depend uniquely
on the relative distance $r$,
$\gamma_i=\gamma_i(r)$.
The modulation of such hopping terms induced
by the lattice displacement determines thus
the electron-phonon interaction.
In full generality, we define thus 
a electron-phonon coupling 
as $\alpha_i=-d|\gamma_i|/dr|_{r=b_i}$
Note that, since the amplitude of the
hopping parameters $|\gamma_i(r)|$ generally decreases with increasing
the distance $r$, we have introduce
an explicit sign (-) in the definition of
$\alpha_i$ so that the corresponding electron-phonon
couplings is {\em by definition} chosen to be positive.

\begin{figure}[t]
\includegraphics[width=8cm,clip=]{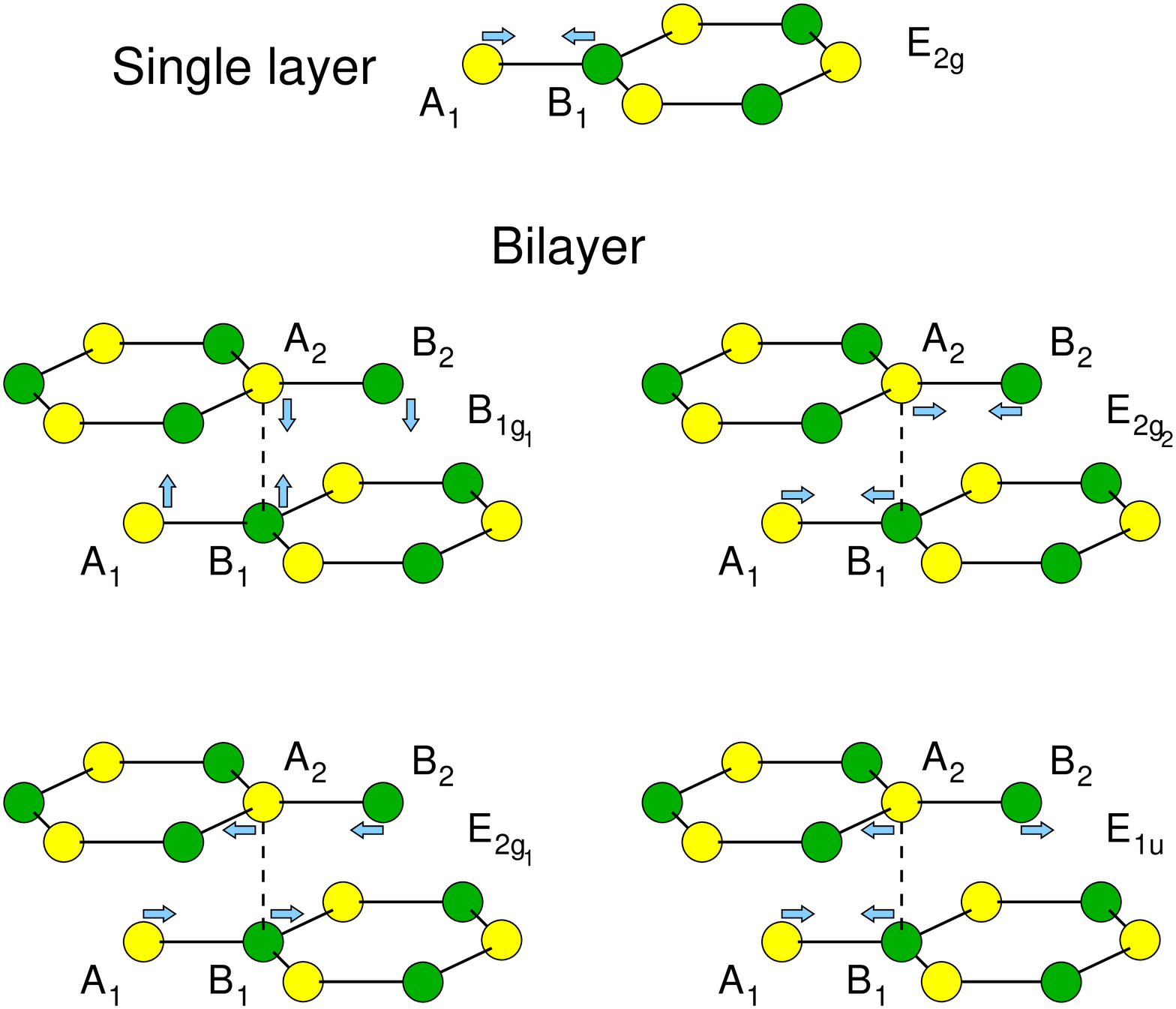}
\includegraphics[width=8cm,clip=]{f-bands.eps}
\caption{(color online) (a) Sketch of atomic displacements
for the relevant phonon lattice modes 
here considered in single-layer graphene ($E_{2g}$),
and in bilayer graphene ($B_{1g_1}$, $E_{2g_2}$, $E_{2g_1}$,
$E_{1u}$).
(b) Representative band structure of single layer and bilayer graphene
upon a frozen phonon lattice distortion (red dashed lines)
with $u=0.1a=0.0246$ \AA\ for the $E_{2g}$ mode
in single layer graphene (left panel)
and for the $E_{2g_2}$ mode in bilayer graphene (right panel).
Also shown is the undistorted band structure
and the band labels ($\epsilon_1$-$\epsilon_4$ from
top to bottom band.}
\label{f-modes}
\end{figure}

In order to reveal the electron-phonon coupling $\alpha_i$
for each hopping parameter $\gamma_i$,
we consider the $E_{2g}$ mode for the single layer
graphene, and the $B_{1g_1}$, $E_{2g_2}$ and $E_{2g_1}$,
for the bilayer graphene as sketched in Fig. \ref{f-modes}a.
We focus on the deformation potential
close to the K point, where one-particle low-energy excitations
are involved, which makes a DFT approach particularly efficient.
We compute the electronic band structure in the presence
of a static frozen phonon displacement by using a plane-wave
implementation\cite{espresso} of the
density functional theory in the local-density approximation (LDA)
for the exchange-correlation potential.\cite{pz}
Ultra-soft pseudopotential for carbon was used with plane-wave
(charge density) cutoff of 40 (400) Ry. 
A uniform wave-vector grid of 18$\times$18 in the irreducible
Brillouin-zone
with cold-smearing of 0.02 Ry was sufficient to converge the calculated quantities to the required accuracy.

In order to provide a common framework for all the
lattice modes of single-layer graphene and well as
of bilayer graphene/graphite, we analyze the deformation
potential as function of $u$, where $u$ represents
the magnitude of the lattice displacement of each atom.
We consider both degenerate in-plane modes along $x$ and $y$ directions,
and the non degenerate out-of-plane modes.
For each case we choose, respectively, ${\bf u}_{{\rm A1},x}=u_x\hat{{\bf x}}$,
${\bf u}_{{\rm A1},y}=u_y\hat{{\bf y}}$,
${\bf u}_{{\rm A1},z}=u_z\hat{{\bf z}}$.
The displacement of the other atoms is thus univocally determined by the 
components of the wavevector of the phonon mode.

Representative electronic structures
of the single-layer and bilayer graphene
in the presence of lattice distortions
are shown in Fig. \ref{f-modes}b.
Focusing at the K point
we can expect, according the different modes considered,
an opening of a gap for the Dirac energy levels
and a further modulation of the upper and lower energy bands.
In the bilayer system,
we label the four $\pi$-bands as $\epsilon_1$-$\epsilon_4$,
from the top to bottom energy, as shown in Fig.  \ref{f-modes}b,
and we denote $\Delta \epsilon$ the possible splitting
of the Dirac state $\Delta \epsilon=\epsilon_2-\epsilon_3$
and $E=\epsilon_1-\epsilon_4$ the energy difference between
the upper and lower band.
We fix for convenience the energy zero 
of our band structure at the Dirac point of the undistorted system.
It is important to stress that our procedure indeed involves
only {\em energy differences} so that the absolute
energy position of the band structure is irrelevant.

It is also useful to introduce here the low-energy Hamiltonian
for the undistorted lattice structures.
Using standard notations,
the single-layer and bilayer graphenes are
thus described respectively by the Hamiltonians:
\begin{eqnarray}
\hat{H}^{\rm 1L}_{\bf k}
&=&
\left(
\begin{array}{cc}
0 & \gamma_0 f_{\bf k} \\
\gamma_0 f_{\bf k}^* & 0
\end{array}
\right),
\label{ham_1L}
\end{eqnarray}
\begin{eqnarray}
\hat{H}^{\rm 2L}_{\bf k}
&=&
\left(
\begin{array}{cccc}
0 & \gamma_0 f_{\bf k}
& \gamma_4 f_{\bf k} & \gamma_3f_{\bf k}^*\\
\gamma_0 f_{\bf k}^*
& \delta & \gamma_1 & \gamma_4f_{\bf k} \\
\gamma_4 f_{\bf k}^* & \gamma_1
& \delta & \gamma_0 f_{\bf k} \\
\gamma_3 f_{\bf k} & \gamma_4 f_{\bf k}^*
& \gamma_0 f_{\bf k}^* & 0 
\end{array}
\right),
\label{ham_2L}
\end{eqnarray}
where
$f_{\bf k}=\mbox{e}^{-ik_xa/\sqrt{3}}+2\mbox{e}^{ik_xa/2\sqrt{3}}\cos(k_ya/2)$,
and where $\delta$ is the difference of the crystal field probed
by the B1-A2 carbon atoms in the bilayer structure with respect to
the A1-B2 atoms.

The band structure for the undistorted bilayer graphene
is shown in Fig. \ref{f-modes}b.
Equating the TB analytical expressions with the computed DFT
eigenvalues  we get  $\epsilon_1=\delta+\gamma_1=0.3620$ eV,
$\epsilon_4=\delta-\gamma_1=-0.3382$ eV, which permits to evaluate
the parameters $\delta=0.0119$ eV and $\gamma_1=0.3501$ eV.

\subsection{Single-layer graphene}

\subsubsection{$E_{2g}$ mode}

With these notations,
we can now consider, as a preliminary check,
the frozen phonon Hamiltonian of the single-layer graphene
upon the $E_{2g}$ distortion.
Along the $x$-direction we have thus:
\begin{eqnarray}
\hat{H}_{\bf k}^{E_{2g}}(u_x)
&=&
\left(
\begin{array}{cc}
0 & \gamma_0 f_{\bf k} +\alpha_0 g_{\bf k} u_x\\
\gamma_0 f_{\bf k}^* +\alpha_0 g_{\bf k}^* u_x& 0
\end{array}
\right),
\label{ham_1L_E2g}
\end{eqnarray}
where
$g_{\bf  k}=2\mbox{e}^{-ik_xa/\sqrt{3}}-2\mbox{e}^{ik_xa/2\sqrt{3}}\cos(k_ya/2)$.
At the K$=4\pi/3a(0,1)$ point, we get $f_{\rm  K}=0$, $g_{\rm K}=3$,
so that
\begin{eqnarray}
\hat{H}_{\rm K}^{E_{2g}}(u_x)
&=&
\left(
\begin{array}{cc}
0 & \Delta_{0,x}  \\
\Delta_{0,x} & 0
\end{array}
\right),
\end{eqnarray}
where $\Delta_{0,x}=3\alpha_0 u_x$.
The degenerate levels at the Dirac point result thus splitted
in single layer graphene upon a $E_{2g}$ lattice distortion
along the $x$-axis
of a total amount $\Delta \epsilon^{E_{2g}}(u)=2|\Delta_{0,x}|
=6\alpha_0 |u_x|$,
in agreement with Refs. \cite{ando1,samsonidze,lazzeri}.

A similar result can be obtained by considering lattice displacements
along the $y$-direction.
In this case we have:
\begin{eqnarray}
\hat{H}_{\bf k}^{E_{2g}}(u_y)
&=&
\left(
\begin{array}{cc}
0 & \gamma_0 f_{\bf k} +\alpha_0 h_{\bf k} u_y\\
\gamma_0 f_{\bf k}^* +\alpha_0 h_{\bf k}^* u_y& 0
\end{array}
\right),
\end{eqnarray}
where
$h_{\bf  k}=i2\sqrt{3}\mbox{e}^{ik_xa/2\sqrt{3}}\sin(k_ya/2)$.
At the K$=4\pi/3a(0,1)$ point, we get $h_{\rm K}=3i$,
so that
\begin{eqnarray}
\hat{H}_{\rm K}^{E_{2g}}(u_y)
&=&
\left(
\begin{array}{cc}
0 & i\Delta_{0,y}  \\
-i\Delta_{0,y} & 0
\end{array}
\right),
\end{eqnarray}
with $\Delta \epsilon^{E_{2g}}(u)=2|\Delta_{0,y}|
=6\alpha_0 |u_y|$ also in this case,
reflecting the perfect degeneracy of the $x$ vs $y$ in-plane lattice vibrations.

\begin{figure}[t]
\includegraphics[width=8cm,clip=]{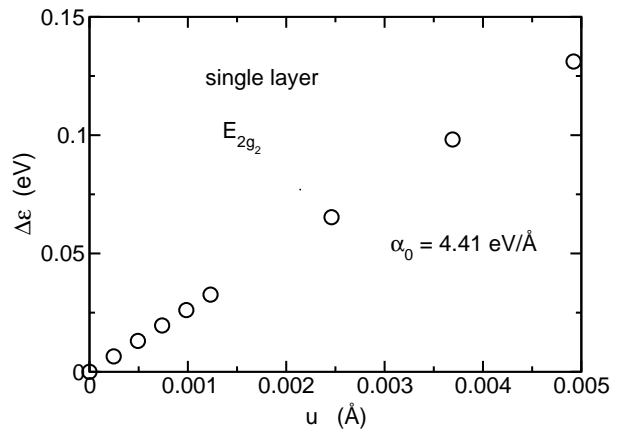}
\caption{Splitting of the Dirac point $\Delta \epsilon$ as a  function of
the $E_{2g}$ lattice distortion ($u$) in single layer graphene as evaluated
from frozen phonon DFT calculations.
The slope of $\Delta \epsilon$ vs. $u$ gives
$\alpha_0=4.41$ eV/\AA.}
\label{f-1L-e2g}
\end{figure}

Our DFT computed splitting is shown in Fig. \ref{f-1L-e2g},
from which we get $\alpha_0=4.41$ eV/\AA, in nice agreement with
Refs. \onlinecite{samsonidze,lazzeri}. We found virtually no difference for
lattice distortions along $x$ or $y$ direction on this
range of $u$.

\subsection{Bilayer graphene}

Once evaluated the in-plane electron-phonon coupling $\alpha_0$
associated with the $\gamma_0$ hopping term in the single-layer
graphene,
we can now address the role of higher order hopping terms in
multilayer graphenes, using the bilayer graphene as a suitable tool.

\subsubsection{$B_{1g_1}$ mode}

\begin{figure}[t]
\includegraphics[width=8cm,clip=]{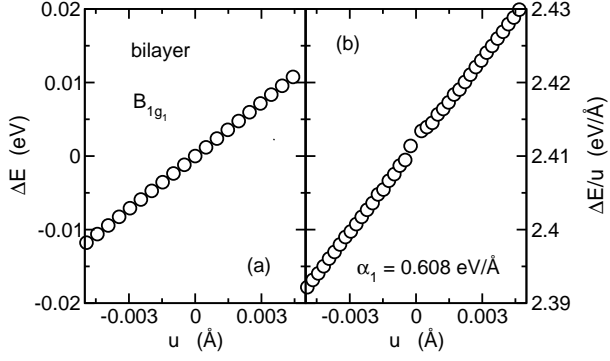}
\caption{(a) Energy difference $\Delta E^{B_{1g_1}}(u)$
between upper and lower bands
as a  function of
the $B_{1g_1}$ lattice displacement $u_z$ in bilayer graphene.
(b) Corresponding ratio $\Delta E^{B_{1g_1}}/u$
as a  function of the lattice distortion $u$.
The extrapolation  for $u \rightarrow 0$
gives $\alpha_1=0.608$ eV/\AA. }
\label{f-B1g1}
\end{figure}

We first consider the out-of-plane $B_{1g_1}$ mode, as depicted
in Fig. \ref{f-modes}a.
This is a quite peculiar mode since it does not lift any symmetry
of the crystal.
We can thus still write the four
energy levels at the K point as
\begin{eqnarray}
\epsilon_{2/3}
&=&
0,
\\
\epsilon_1
&=&
\delta(u)+\gamma_1(u),
\\
\epsilon_4
&=&
\delta(u)-\gamma_1(u),
\end{eqnarray}
where we have explicitly expressed the intrinsic
dependence of the parameters $\delta$ and $\gamma_1$
on the $B_{1g_1}$ lattice distortion.
We can note that,
as a consequence of the symmetry preserving displacements,
no gap is opened at the K point.
Useful information is however encoded in
the energy difference between the high-energy bands
$E^{B_{1g_1}}(u)=\epsilon_1(u)-\epsilon_4(u)$
 which, from Hamiltonian\ref{ham_2L}, results
\begin{eqnarray}
E^{B_{1g_1}}(u)
&=&
2\gamma_1
+4\alpha_1u.
\end{eqnarray}
We can evaluate thus the electron-phonon coupling
$\alpha_1$ from the linear dependence
of $\Delta E^{B_{1g_1}}(u)=[E^{B_{1g_1}}(u)-E^{B_{1g_1}}(0)]$ on $u$.
The calculated  DFT dependence of $\Delta E^{B_{1g_1}}(u)$
as a function of the vertical displacement $u_z$ is shown in
Fig. \ref{f-B1g1}a, whereas the ratio $\Delta E^{B_{1g_1}}(u)/u$
is shown in Fig. \ref{f-B1g1}b, whose extrapolation for
$u_z\rightarrow 0$ gives $\alpha_1=0.608$ eV/\AA.

\subsubsection{$E_{2g_2}$ mode}
\label{sse2g2}

The $B_{1g_1}$ mode is quite peculiar as, since it does not lift any
symmetry of the system, it does not split the Dirac energy levels
at the K point. We have shown above however that the splitting
of the additional upper and lower bands can be used
to estimate the electron-phonon coupling associated
with $\gamma_1$. Things are richer when other modes,
reducing the symmetry of the crystal, are considered.
In this case useful information about {\em different}
electron-phonon coupling are encoded in the
splitting of the Dirac point as well as in the $u$-dependence
of the differences between high-energy bands, $\Delta E(u)$.

Let us consider for instance
the electronic structure of the bilayer graphene
under a $E_{2g_2}$ lattice distortion.
If we consider only the leading order
linear electron-phonon couplings
$\alpha_i$,
we can thus write
\begin{widetext}
\begin{eqnarray}
\hat{H}_{\bf k}^{E_{2g_2}}(u_x)
&=&
\left(
\begin{array}{cccc}
0 & \gamma_0 f_{\bf k}+\alpha_0 g_{\bf k} u_x
& \gamma_4f_{\bf k} & \gamma_3f_{\bf k}^* -\cos\theta \alpha_3 g_{\bf k}^* u_x\\
\gamma_0 f_{\bf k}^* +\alpha_0 g_{\bf k}^* u_x 
& \delta & \gamma_1 & \gamma_4f_{\bf k} \\
\gamma_4f_{\bf k}^* & \gamma_1
& \delta & \gamma_0 f_{\bf k} +\alpha_0 g_{\bf k} u_x \\
\gamma_3f_{\bf k} -\cos\theta\alpha_3 g_{\bf k} u_x& \gamma_4f_{\bf k}^*
& \gamma_0 f_{\bf k}^* +\alpha_0 g_{\bf k}^* u_x& 0 
\end{array}
\right),
\label{ham_2L_Eg}
\end{eqnarray}
\end{widetext}
where $\cos\theta=b/\sqrt{b^2+d^2}\approx 0.39$
is a geometric factor accounting for the
projection of the lattice displacement along the direction of
the $\gamma_3$ bond.
Evaluated at the K point, it reads 
\begin{eqnarray}
\hat{H}_{\rm K}^{E_{2g_2}}
&=&
\left(
\begin{array}{cccc}
0 &  \Delta_{0,x} &  0 & -\Delta_{3,x} \\
\Delta_{0,x} & \delta & \gamma_1 &   0 \\
0 & \gamma_1 & \delta &  \Delta_{0,x}  \\
-\Delta_{3,x} & 0 &  \Delta_{0,x}  & 0
\end{array}
\right),
\label{ham_2Lux}
\end{eqnarray}
where 
$\Delta_{3,x/y}=3\cos\theta \alpha_3 u_{x/y}$.

\begin{figure}[t]
\includegraphics[width=8cm,clip=]{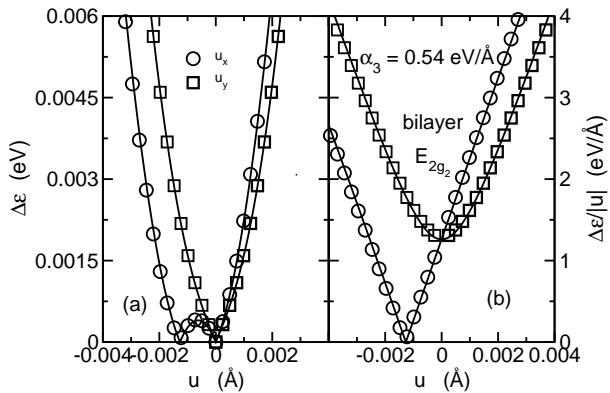}
\caption{(a) Splitting of the Dirac point
$\Delta \epsilon$ as a  function of
the $E_{2g_2}$ lattice distortion $u$ in bilayer graphene.
Empty circles are DFT frozen phonon calculations for $u_x$,
while empty squares for $u_y$.
Solid lines are the solution of the corresponding analytical models
in Eqs. (\ref{secularEgx}), (\ref{secularEgy}), using
$\delta=0.0119$ eV, $\gamma_1=0.35$ eV and
$\alpha_3=0.54$ eV/\AA. The value of
$\alpha_0$ eV/\AA\, is irrelevant on this quantity in this range.
(b) Corresponding $\Delta \epsilon/|u|$
as a  function of the lattice distortion $u$.
The extrapolation of the DFT data for $u \rightarrow 0$
gives an unbiased estimate of $\alpha_3=0.54$ eV/\AA.}
\label{f-eg_2L}
\end{figure}

The eigenvalues $\epsilon$'s
can be thus obtained from the secular equation:
\begin{eqnarray}
&&\epsilon^2(\delta-\epsilon)^2
+2\Delta_{0,x}^2\epsilon (\delta-\epsilon)
-\epsilon^2\gamma_1^2
+2\Delta_{0,x}^2\Delta_{3,x}\gamma_1
\nonumber\\
&&+\Delta_{0,x}^4
-\Delta_{3,x}^2(\delta-\epsilon)^2
+\Delta_{3,x}^2\gamma_1^2
=
0,
\label{secularEgx}
\end{eqnarray}
Eq. (\ref{secularEgx}) predicts  a linear splitting
of the Dirac levels as a function of $u_x$.
Linearizing with respect to $u_x$ we find:
\begin{eqnarray}
\Delta \epsilon^{E_{2g_2}}=2|\Delta_{3,x}|
=6\cos\theta\alpha_3 |u_x|,
\end{eqnarray}
which permits us to evaluate $\alpha_3$ from 
the linear splitting at the K point of the Dirac bands in 
bilayer graphene upon a $E_{2g_2}$ lattice distortion. 
In Fig. \ref{f-eg_2L}a we show the linear splitting
$\Delta \epsilon_{\rm K}^{E_{2g_2}}$ computed by using
our frozen phonon DFT calculations for different $u_x$  (open circles).
The linear extrapolation of $\Delta \epsilon_{\rm K}^{E_{2g_2}}/u_x$
for d$u_x \rightarrow 0$, as shown in Fig. \ref{f-eg_2L}b,
gives us thus an unbiased estimate of $\alpha_3=0.54$ eV/\AA.

The accuracy of such estimate, as well as of the
tight-binding analysis here considered, can be checked
by using this last value ($\alpha_3=0.54$ eV/\AA) and the
TB parameters previously evaluated in an independent way
in the undistorted structure ($\delta=0.0119$ eV, $\gamma_1=0.35$ eV)
to calculate the splitting on a wider range of $u_x$,
without the linearization, but solving  Eq. (\ref{secularEgx}).
The analytical results obtained in this way are in excellent 
agreement with DFT calculations proving thus the full intrinsic
consistency of the value of $\alpha_3$ with respect 
the other TB parameters.

Note also that the DFT calculations 
predict a critical value $\bar{u}_x$ where the gap
at the K point close, reconstructing there thus,
for this particular value of $u_x$, a Dirac cone.
This peculiar feature can also be understood
using the TB model.
As a matter of fact, from an inspection of
Eq. (\ref{secularEgx}), one can find two very close
critical values $\bar{u}_x=
-\alpha_3\cos\theta(\gamma_1\pm\delta)/3\alpha_0^2$
where the gap at the K point closes.
These points are however so close that they cannot
be resolved on the scale of Fig. \ref{f-eg_2L}.
The reconstruction of the Dirac cone at the K point
is a mixed combination of the effects of the trigonal warping 
induced by $\gamma_3$ and of the additional effects
related to the lattice distortion.
In the undistorted structure, indeed, we know 
that the effect of $\gamma_3$ in bilayer systems is to induce satellite
Dirac cones at finite ${\bf k}$ in addition to the
conventional one at the K point.
Lattice distortions induce, as well as in single-layer graphene,
a shift of the main Dirac point away from the K point,
opening thus there a gap.
The satellite Dirac points however move {\em as well} as functions
of the lattice distortion.  At a certain value, $\bar{u}_x$,
one of the satellite Dirac points is moved again across the K point,
and this feature is reflected in the closing of the gap
in Fig. \ref{f-eg_2L} at a finite $u_x$. The value
of $\bar{u}_x$ agrees also in excellent way with
the above analytical estimate from the tight-binding model.
On the other hand, for $u_y$ displacements, the Dirac point
moves in an orthogonal direction with respect to the K point
and no reconstruction of Dirac cones at K is possible.
A more detailed analysis of this issue is provided
in Appendix \ref{app-cone}.

Finally, as a last check of our analysis,
we computed also the frozen phonon
energy splitting for $E_{2g_2}$ lattice displacements along $y$.
DFT calculations are shown in Fig. \ref{f-eg_2L} as empty squares.
To extract information about the electron-phonon coupling,
we analyze the Hamiltonian at the K point which reads now:
\begin{eqnarray}
\hat{H}_{\rm K}^{E_{2g_2}}
&=&
\left(
\begin{array}{cccc}
0 &  -i\Delta_{0,y} &  0 &- i\Delta_{3,y} \\
i\Delta_{0,y} & \delta & \gamma_1 &   0 \\
0 & \gamma_1 & \delta & - i\Delta_{0,y}  \\
i\Delta_{3,y} & 0 &  i\Delta_{0,y}  & 0
\end{array}
\right),
\label{ham_2Luy}
\end{eqnarray}
with a secular equation:
\begin{eqnarray}
&&\epsilon^2(\delta-\epsilon)^2
+2\Delta_{0,y}^2\epsilon (\delta-\epsilon)
-\epsilon^2\gamma_1^2
\nonumber\\
&&
+\Delta_{0,y}^4
-\Delta_{3,y}^2(\delta-\epsilon)^2
+\Delta_{3,y}^2\gamma_1^2
=0.
\label{secularEgy}
\end{eqnarray}
Note that, unlike the displacements along $x$
[Eq. (\ref{secularEgx})],
Eq. (\ref{secularEgy}) is symmetric with respect
to $u_y \rightarrow -u_y$.
For small values of $u_y$, we once more obtain
\begin{eqnarray}
\Delta \epsilon^{E_{2g_2}}=2|\Delta_{3,y}|
=6\cos\theta\alpha_3 |u_y|,
\end{eqnarray}
reflecting the degeneracy, at the linear level,
of the $E_{2g_2}$ mode along the two directions.
The extrapolation of $\Delta\epsilon/u_y$ coincides
with $\Delta\epsilon/u_x$ for $u \rightarrow 0$,
providing thus the same value
$\alpha_3=0.54$ eV/\AA.

It is also interesting to give a look now  at the
dependence of  $E^{2g_2}$ 
at the K point with respect to the lattice displacement $u_x$.
For these levels we find a quadratic dependence on $u_x$.
Expanding Eq. (\ref{secularEgy})
at the second order with respect to $u_x$,
we get
\begin{eqnarray}
\epsilon_1
&=&
\delta+\gamma_1
+\frac{\Delta_{0,x}^2}{\gamma_1+\delta},
\end{eqnarray}
\begin{eqnarray}
\epsilon_4
&=&
\delta-\gamma_1
-\frac{\Delta_{0,x}^2}{\gamma_1-\delta},
\end{eqnarray}
so that
\begin{eqnarray}
\Delta E^{2g_2}
&=&
2\gamma_1
+\frac{2\gamma_1\Delta_{0,x}^2}{\gamma_1^2-\delta^2}.
\label{Esplit}
\end{eqnarray}

\begin{figure}[t]
\includegraphics[width=8cm,clip=]{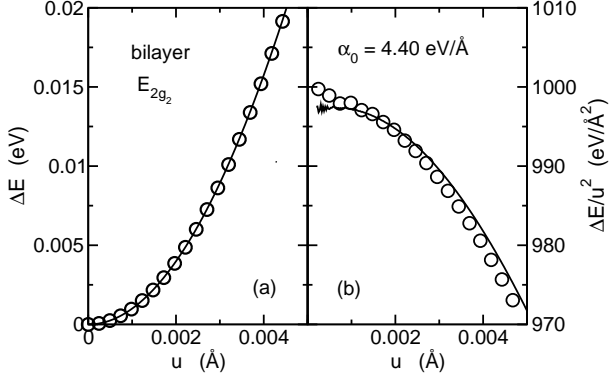}
\caption{(a) Variation of $\Delta E$ as a  function of
the $E_{2g_2}$ lattice distortion $u_x$ in bilayer graphene.
(b) Corresponding ratio $\Delta E/u^2$.
The extrapolation for $u_x \rightarrow 0$ gives
$\alpha_0=4.40$ eV/\AA.
Empty circles are DFT data, while
solid lines are obtained from the TB model
with 
$\delta=0.0119$ eV, $\gamma_1=0.35$ eV and
$\alpha_0=4.40$ eV/\AA\, and $\alpha_1=0.68$  eV/\AA.
The value of $\alpha_3$ eV/\AA\, is irrelevant on this quantity
in this range.}
\label{f-e2g2_2L}
\end{figure}

The DFT calculations (open symbols)
of the $u$-dependence of 
$\Delta E^{2g_2}$ are shown in  Fig. \ref{f-e2g2_2L}
(panel a), as well with the ratio 
$\Delta E^{2g_2}/u_x^2$ (panel b).
The extrapolation of $\Delta E^{2g_2}/u_x^2$ for $u_x \rightarrow 0$
provides thus an estimate  $\alpha_0=4.40$ eV/\AA\ 
which essentially coincides with the value extracted
in the single-layer graphene.

It should be here noted that Eq. (\ref{Esplit}) has been derived
from Eq. (\ref{ham_2L_Eg}) where only the linear terms in $u$ where
retained. 
Some care is however needed on this regards since
we are actually investigating here a {\em quadratic}
dependence on $u$.
A careful analysis shows that
further corrections at the quadratic order in Eq. (\ref{Esplit})
appear through the explicitly dependence 
of $\gamma_1$ on $u$. Taking into account the geometry of the lattice
displacement, one should write thus
\begin{eqnarray}
\Delta E^{2g_2}
&=&
2\gamma_1-\frac{4\alpha_1 u_x^2}{c}
+\frac{2\gamma_1\Delta_{0,x}^2}{\gamma_1^2-\delta^2}.
\label{Esplit2}
\end{eqnarray}
The correction coming from $\alpha_1$ are however
two orders of magnitude smaller that the term $\propto \Delta_{0,x}$
and they are here ineffective.

\subsubsection{$E_{2g_1}$ mode}

After having determined the electron-phonon coupling
$\alpha_0$, $\alpha_1$, $\alpha_3$ in bilayer graphene
from the frozen phonon dependence of the energy levels
at the K point under $B_{1g_1}$ and $E_{2g_2}$ distortions,
we are now aiming to a corresponding characterization
of the last remaining parameter $\alpha_4$ associated
with the $\gamma_4$ hopping.
The most straightforward way to probe it,
as we are going to see, is to consider the $E_{2g_1}$ phonon mode,
as depicted in Fig. \ref{f-modes}.

Upon distortion along the E$_{2g_1}$ phonon mode, the Hamiltonian reads: 

\begin{widetext}
\begin{eqnarray}
\hat{H}_{\rm K}^{E_{2g_1}}(u_x)
&=&
\left(
\begin{array}{cccc}
0 & \gamma_0 f_{\bf k} 
& \gamma_4f_{\bf k} +\cos\theta\alpha_4 g_{\bf k} u_x
& \gamma_3f_{\bf k}^* -\cos\theta \alpha_3 g_{\bf k}^* u_x\\
\gamma_0 f_{\bf k}^* & \delta & \gamma_1 &
\gamma_4f_{\bf k} +\cos\theta\alpha_4 g_{\bf k} u_x\\
\gamma_4f_{\bf k}^* +\cos\theta\alpha_4 g_{\bf k}^* u_x & \gamma_1
& \delta & \gamma_0 f_{\bf k}  \\
\cos\theta \gamma_3f_{\bf k} -\alpha_3 g_{\bf k} u_x&
\gamma_4f_{\bf k}^* +\cos\theta\alpha_4 g_{\bf k}^* u_x
& \gamma_0 f_{\bf k}^* & 0 
\end{array}
\right).
\label{ham_2L_Eg1}
\end{eqnarray}
\end{widetext}
Evaluated at the K point, we thus have:
\begin{eqnarray}
\hat{H}_{\rm K}^{E_{2g_1}}
&=&
\left(
\begin{array}{cccc}
0 &  0 &  \Delta_{4,x} & -\Delta_{3,x} \\
0 & \delta & \gamma_1 &   \Delta_{4,x} \\
\Delta_{4,x} & \gamma_1 & \delta &  0  \\
-\Delta_{3,x} &  \Delta_{4,x}&  0  & 0
\end{array}
\right),
\label{ham_2Leg1x}
\end{eqnarray}
where $\Delta_{4,x}=3\cos\theta\alpha_4 u_{x}$,
and we can write the secular equation:
\begin{eqnarray}
&&\epsilon^2(\delta-\epsilon)^2
+2\cos^2\theta\Delta_{4,x}^2\epsilon (\delta-\epsilon)
-\epsilon^2\gamma_1^2
\nonumber\\
&&+2\cos^3\theta\Delta_{4,x}^2\Delta_{3,x}\gamma_1
+\cos^4\theta\Delta_{4,x}^4
-\cos^2\theta\Delta_{3,x}^2(\delta-\epsilon)^2
\nonumber\\
&&+\cos^2\theta\Delta_{3,x}^2\gamma_1^2
=0.
\label{secularEg1x}
\end{eqnarray}

Eq. (\ref{secularEg1x}), predicts also, as
(\ref{secularEgx}), a linear splitting of the Dirac level
upon lattice distortion associated once more with
$\alpha_3$, i.e.
$\Delta \epsilon_{\rm K}^{E_{2g_1}}
=6\cos\theta\alpha_3 |u_x|$.
The value of $\alpha_3$ estimated upon
such lattice distortion coincides with the one
obtained previously using the  $E_{2g_2}$ mode,
corroborating thus the analysis.

More useful information is however encoded
in the frozen phonon dependence of $\Delta E$.
Such splitting was
above employed to estimate directly $\alpha_0$ from the
frozen phonon $E_{2g_2}$ lattice distortion.
In the present $E_{2g_1}$ context, we can see that
we still get, although not a direct, an indirect estimate
of $\alpha_4$ from the $u$-dependence of $\Delta E$.
We can indeed write
\begin{eqnarray}
\epsilon_1
&=&
\delta+\gamma_1
+\frac{\Delta_{4,x}^2}{\gamma_1+\delta},
\end{eqnarray}
and
\begin{eqnarray}
\epsilon_4
&=&
\delta-\gamma_1
-\frac{\Delta_{4,x}^2}{\gamma_1-\delta},
\end{eqnarray}
so that $\Delta E$ is expected once more
to presents a quadratic dependence on $u$.
Taking into account, just as in the $E_{2g_2}$ case,
the quadratic dependence associated with $\gamma_1$,
we can write thus
\begin{eqnarray}
\Delta E^{2g_1}
&=&
2\gamma_1-\frac{4\alpha_1 u_x^2}{c}
+\frac{2\gamma_1\Delta_{4,x}^2}{\gamma_1^2-\delta^2}.
\label{Esplit3}
\end{eqnarray}

\begin{figure}[t]
\includegraphics[width=8cm,clip=]{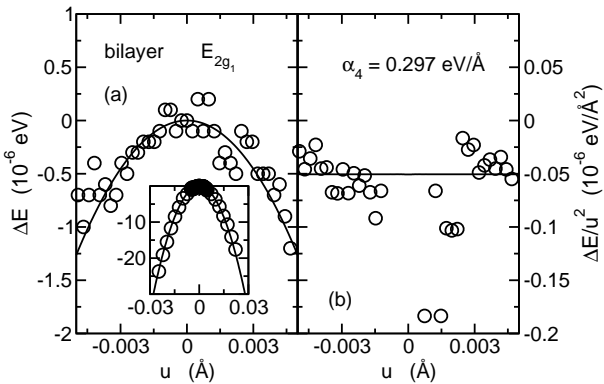}
\caption{(a) $\Delta E$ as a  function of
the $E_{2g_1}$ lattice distortion $u_x$ in bilayer graphene.
Inset: same on a wider $u$-region. Axis labels in the inset
are the same as in the main panel.
(b) Corresponding ratio $\Delta E/u_x^2$.
From the extrapolation for $u_x \rightarrow 0$, and
taking into account the contribution of the
$u$ dependence of $\gamma_1$, we can estimate
$\alpha_4=0.32$ eV/\AA.
Empty circles are DFT data, while
solid lines are obtained from the TB model
with 
$\delta=0.0119$ eV, $\gamma_1=0.35$ eV and
$\alpha_1=0.61$ eV/\AA\, and $\alpha_4=0.30$  eV/\AA.
The value of $\alpha_0$ eV/\AA\ is irrelevant on this quantity
in this range.
}
\label{f-e2g1}
\end{figure}

DFT calculations
for $\Delta E^{2g_1}$
are shown in Fig. \ref{f-e2g1}a on the same
$u$-scale employed for other lattice modes.
Due to the smallness of such $u$-dependence,
numerical noise is here much larger than
in previous analyses.
A negative quadratic curvature can be however still clearly observed,
which is better visible in a larger $u$-window in the inset.
Such negative curvature is at odds with the
$u$-dependence of $\Delta E^{2g_1}$ coming from
the contribution alone of $\alpha_4$ as predicted in
Eq. (\ref{Esplit3}).
This suggests that the negative contribution from $\gamma_1$
is here of the same order of the term $\propto \Delta_{4,x}^2$.
On the other hand, the $\alpha_1$ term alone
would give an extrapolation of the ratio $\Delta E/u_x^2$
at $u_x \rightarrow 0$ of the order
$\lim_{u_x \rightarrow 0} \Delta E/u_x^2 \approx -0.73$ eV/\AA$^2$
much bigger than what observed
As a matter of fact, we can nicely reproduce the DFT data
by taking $\alpha_4=0.30$  eV/\AA.
The comparison between DFT calculations and the TB model
with this value of $\alpha_4$ reasonably good, as shown
in Fig. \ref{f-e2g1}. We have to stress however that, unlikely
the other parameters $\alpha_i$ that were obtained
in a direct unbiased way by a
high-precision extrapolation for $u \rightarrow 0$,
since $\alpha_4$ was deducted in an indirect way from 
the knowledge of $\alpha_1$, and given the numerically
scattered DFT data in Fig. \ref{f-e2g1}, this value
$\alpha_4=0.30$  eV/\AA\, must be considered just
as an indicative electron-phonon coupling for
this hopping parameter.

\subsubsection{Other modes ($E_{1u}$, $B_{1g_2}$, \ldots)}

Other optical modes at ${\bf q}=0$ can be in principle considered
to investigate the deformation potential due
the electron-phonon interaction. However, they result 
to be not particularly convenient in order to disentangle the role
of the different electron-phonon couplings associated
with the different hopping parameters.
Once can see for instance that the $E_{1u}$ (also shown
in Fig. \ref{f-modes}) induces a {\em quadratic} splitting
of the Dirac point as a function of $u$,
whose curvature depends on the same level
on both $\alpha_0$ and $\alpha_4$, so that their values
cannot be estimated in an unbiased way from an extrapolation
for $u \rightarrow 0$. Similar problems appear when
considering the splitting of high energy bands
for $E_{1u}$, or the energy splitting (Dirac point as well as
high-energy bands) for the other modes.
Also in these cases, the deformation potential results to be
a mixing of different electron-phonon coupling, making
the quantitative evaluation of the $\alpha_i$ from these modes
not reliable. We have however checked, on the other hand,
that the above values estimated from the $B_{1g_1}$
$E_{2g_2}$ and $E_{2g_1}$ modes reproduce the energy differences of the electronic bands at the K point 
upon other different lattice modes.

\section{Discussion and conclusions}

\begin{figure}[t]
\includegraphics[width=8cm,clip=]{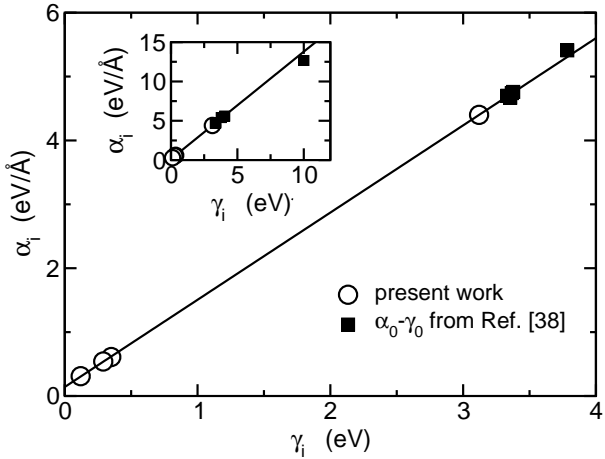}
\caption{Plot of the $\alpha_i$ vs. $\gamma_i$
parameters obtained from different approaches.
Empty circles are data obtained by the present work
where $\alpha_i$ was estimated by the frozen-phonon technique
and $\gamma_i$, when not available, were taken 
from Ref. \onlinecite{pp}. Filled squares are data collected
by Ref. \onlinecite{lazzeri} using a wide variety
of techniques, including Hartree-Fock, the hybrid B2LYP
functional, LDA, GGA and GW. Inset: same data on a larger scale.}
\label{f-corr}
\end{figure}

In this paper we have employed a combined TB and DFT approach
to evaluate the deformation potential in single-layer and
bilayer graphene associated with the modulation
of the different hopping parameters.
In order to avoid any fitting procedure,
we have focused on the low-energy levels $\epsilon_\nu$ at the
high-symmetry point K and we have characterized
the electron-phonon coupling $\alpha_i$ for each
hopping term by a careful analysis of the
frozen-phonon dependence of $\epsilon_\nu$
upon the lattice displacement for different lattice modes.
In this way we were able to determine within
a unique framework all the deformation
potentials $\alpha_i$ for both the intralayer ($i=0$) and
interlayer hoppings ($i=1,3,4$) as well as
the TB parameters $\gamma_1$, $\delta$.
We summarize in Table\ref{t-param} our results
for $\alpha_i$.
\begin{table}[t]
\begin{center}
\begin{tabular}{ccc}
\hline \hline
$i$ &\hspace{3mm} $\alpha_i$ (eV/\AA)\hspace{3mm} & $|\gamma_i|$ (eV) \\
\hline 
0 (1L) & 4.41 & 3.12$^*$ \\
0 (2L) & 4.40 & 3.12$^*$ \\
1 (2L) & 0.61 & 0.35$^\dagger$ \\
3 (2L) & 0.54 & 0.29$^*$ \\
4 (2L) & 0.30 & 0.12$^*$
\\
\hline
\multicolumn{3}{l}{$^*$ From Ref. \cite{pp}} \\
\multicolumn{3}{l}{$^\dagger$ present work}
\end{tabular}
\end{center}
\caption{Electron-phonon coupling $\alpha_i$
associated with each hopping parameter $\gamma_i$
in single layer (1L) and bilayer (2L) graphene.
We also show, in the right column, representative values of the
TB parameters $\gamma_1$. We provide an  estimate
of $\gamma_1$, while $\gamma_i$ for $i=0,3,4$ are taken
from Ref. \onlinecite{pp}.}
\label{t-param}
\end{table}
We can also compare these values with the
estimates of the absolute magnitude of the corresponding
hopping parameters, as reported in the right column in Table
\ref{t-param}. The correlation between these two quantities
is also shown in Fig. \ref{f-corr} which reveals an almost perfect
linear scaling of $\alpha_i$ with $\gamma_i$.
A mean-square fitting procedure gives
\begin{equation}
\alpha_i
=
A+B\gamma_i
\end{equation}
where $A=0.141$ eV/\AA\, and $B=1.365$ \AA$^{-1}$.
We would like to stress the importance of
such robust
underlying correlation between the magnitude of the hopping term
and the corresponding electron-phonon interaction
{\em independently} on the precise value of $\gamma_i$.
It is indeed well known that
the estimates of the hopping parameters $\gamma_i$
can significantly depend on the fitting procedure
as well as on the inclusion of
many-body effects in first-principles band structure for example, within the GW scheme.
A detailed study of this issue,
including also Hartree-Fock (HF) calculations,
is provided in
Ref. \onlinecite{gruneis,lazzeri}, where they also estimate
{\em within the same level of approximation}
the overall electronic $\pi$-bandwidth, related to $\gamma_0$,
and the electron-phonon coupling $\alpha_0$
in single-layer graphene and graphite.
Their results are also plotted in Fig. \ref{f-corr}, where
we have translated the high-energy $\pi$-band splitting
$\Delta \epsilon_{\rm M}$ at the M point
in the hopping parameter through the phenomenological
relation $\Delta \epsilon_{\rm M}=1.21 \gamma_0$.
Also in this case, considering the widest variety
of approaches (HF, LDA, GGA, hybrid B3LYP functional
and GW), the trend is almost perfectly linear.

Apart the fundamental implications of this result, it suggests
a well, defined way to estimate {\em experimentally}
the size of the electron-phonon coupling
once the band parameters $\gamma_i$ are extracted
experimentally, for instance by means angle-resolved
photoemission spectroscopy (ARPES).
In particular, the evolution of the electron-phonon coupling can be followed as a function of doping, applied 
electric-field, strain, etc...
This can be can in a quite easy and safe way for $\gamma_0$,
by looking at the linear conical dispersion at the K point,
and for $\gamma_1$, by looking at the upper and lower
band energy splitting at the same K point in bilayer graphene
and graphite.
Experimental determinations of $\gamma_3$ and $\gamma_4$
have been also provided in literature.

Our analysis provides thus a crucial, and previously missing,
information to include quantitatively  the role of the lattice deformations
on the electronic, transport and optical properties of multilayered graphene.
The effects of the lattice deformations on the electronic structure can be included in TB models involving
the deformation potential associated
with higher hopping terms than the nearest-neighbor ones.

\acknowledgments
E.C. acknowledges support from the European FP7 Marie
Curie project PIEF-GA-2009-251904 and
G.P. from  CINECA-HPC ISCRA supercomputing grant.

\appendix
\label{app-cone}
\section{Dirac cone reconstruction upon $E_{2g_s}$
lattice distortion}

In this Appendix we discuss in more details the origin and the
phenomenology of the reconstruction of the Dirac point
at the K edge for a critical value of the $E_{2g_2}$ lattice distortion,
as pointed out by DFT calculations in Fig. \ref{f-e2g2_2L}
and confirmed by the TB model.

As a starting point we remind that in realistic undistorted bilayer
graphenes,
electronic processes like
the ``skew'' hopping $\gamma_3$ split the
the parabolic Dirac cone in four linear Dirac points.\cite{mccann}
In the simplest TB model with only $\gamma_0$-$\gamma_1$-$\gamma_3$ hoppings,
the four Dirac points are located respectively at ${\bf k}=(0,0)$,
$(k_3,0)$, $(-k_3/2,\sqrt{3}k_3/2)$, $(-k_3/2,-\sqrt{3}k_3/2)$, 
where $k_3=\gamma_1\gamma_3/\gamma_0\hbar v_{\rm F}$.\cite{mccann}

In order to investigate the role of the $E_{2g_2}$ lattice distortion,
we expand the Hamiltonian (\ref{ham_2L_Eg}) for small but finite
${\bf k}=(k_x,k_y)$.
Neglecting here for simplicity the terms $\gamma_4$, $\delta$ that
break the particle-hole symmetry,
we can thus write:
\begin{eqnarray}
\hat{H}_{\bf k}^{E_{2g_2}}(u_x)
&=&
\hbar v_{\rm F}\left(
\begin{array}{cccc}
0 & \pi_{0,u}
& 0 & v_3 \pi_{3,u }^*\\
\pi^*_{0,u}
& 0 & \tilde{\gamma}_1 & 0 \\
0 & \tilde{\gamma}_1
& 0 &   \pi_{0,u}\\
 v_3 \pi_{3,u}& 0
&  \pi^*_{0,u}& 0 
\end{array}
\right),
\end{eqnarray}
where $\pi_{0,u}=k_x+ik_y+a_0u_x$, $\pi_{3,u}=k_x+ik_y-a_3u_x$,
and where $a_0=3\alpha_0/\hbar v_{\rm F}$,
$a_3=3\alpha_3\cos\theta/\hbar v_{\rm F}$, 
$\tilde{\gamma}_1=\gamma_1/\hbar v_{\rm F}$.

In the absence of particle-hole asymmetry, the four Dirac cones
lie at the same energy $\epsilon=0$ also in the presence
of lattice distortion.
We can thus trace their evolution as a function of $u_x$
by analyzing the solution
\begin{equation}
\mbox{det}\left[\hat{H}_{\bf k}^{E_{2g_2}}(u_x)\right]
=
0.
\label{det}
\end{equation}

\begin{figure}[t]
\includegraphics[width=8cm,clip=]{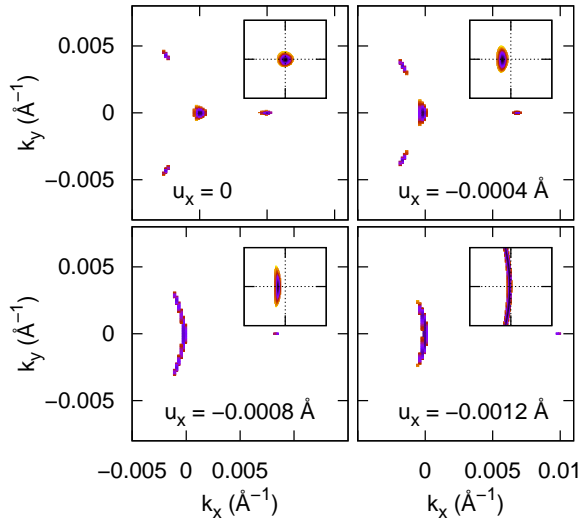}
\caption{(color online) Evolution of the low-energy Dirac-like states
($\epsilon \approx 0$) as a function of the $E_{2g_2}$
lattice displacements $u_x$. Inset: a zoom in the region
$|k_x|, |k_y| \le 0.001$ \AA$^{-1}$. Colors refer here to
the energy distance from the Dirac points at $\epsilon=0$,
The color scale has been adapted
in each panel to make more visible the low energy states, with
$\epsilon=0$ being the darked regions.}
\label{f-dirac}
\end{figure}

The evolution of the Dirac points, corresponding to the low-energy
states of (\ref{det}), as a function of $u_x$,
in the relevant region $u_x < 0$, is shown in
Fig. \ref{f-dirac}. The inset shows also a zoom
close to the K point.
In similar way as it has been reported for uniaxial strain,\cite{mucha}
also upon the optical $E_{2g_2}$ lattice distortion the Dirac points shift away
from their original location for $u=0$. While such shift is
monotonic for the three ``leg parts'', the shift of the central one
is however non monotonic, with a initial departure from
the K point, followed by a turn back along the opposite direction.
Hence, at a critical value $\bar{u}_x=-\tilde{\gamma}_1 a_3/a_0^2$
the ``central part" will eventually cross again the K point
and then continue moving on the opposite side.

We can quantify this evolution by focusing on the axis $k_x$
and tracing the evolution of the roots of Eq. (\ref{det}) for $k_y=0$.
A straightforward analysis gives thus:
\begin{eqnarray}
\bar{k}_{x,\pm}
&=&
\frac{\tilde{\gamma}_1 v_3-2a_0u_x}{2}
\nonumber\\
&&
\pm
\frac{1}{2}
\sqrt{
\tilde{\gamma}_1^2 v_3^2
-4\tilde{\gamma}_1 v_3a_0u_x
-4\tilde{\gamma}_1a_3u_x 
},
\label{npp3}
\end{eqnarray}
where $\bar{k}_{x,-}$ is the non monotonic solution for $u_x<0$
starting from ${\bf k}=(0,0)$ at $u_x=0$
and $\bar{k}_{x,+}$ is the second shifting away solution
starting from ${\bf k}=(k_3,0)$.
From Eq. (\ref{npp3}) we thus get a critical value
$\bar{u}_x=-\tilde{\gamma}_1a_3/a_0^2=
-\alpha_3\cos\theta\gamma_1/3\alpha_0^2$.
Similar calculations can be generalized including
the crystal field $\delta$ which breaks the particle-hole
symmetry.
We get in this case the result
$\bar{u}_x=-\alpha_3\cos\theta(\gamma_1\pm\delta)/3\alpha_0^2$,
as reported in  Sec. \ref{sse2g2}.


\begin{thebibliography}{99}

\bibitem{chen}
J.H. Chen, C. Jang, S. Xiao, M. Ishigami, and M.S. Fuhrer,
Nat. Nanotech. {\bf 3}, 206 (2008).

\bibitem{fratini}
S. Fratini and F. Guinea,
Phys. Rev. B {\bf  77}, 195415 (2008).

\bibitem{hwang}
E.H. Hwang and S. Das Sarma,
Phys. Rev. B {\bf 77}, 115449 (2008).

\bibitem{efetov}
D.K. Efetov and P. Kim,
Phys. Rev. Lett. {\bf 105}, 256805 (2010).

\bibitem{ochoa}
H. Ochoa, E.V. Castro, M.I. Katsnelson, and F. Guinea,
Phys. Rev. B {\bf 83}, 235416 (2011).

\bibitem{voz}
M.A.H. Vozmediano, M.I. Katsnelson, and F. Guinea,
Phys. Rep. {\bf 496}, 109 (2010).

\bibitem{lopes}
J.M.B. Lopes dos Santos, N.M.R. Peres, and A.H. Castro Neto,
Phys. Rev. Lett. {\bf 99}, 256802 (2007); arXiv:1202.1088 (2012).

\bibitem{dejuan}
F. de Juan, A. Cortijo, M.A.H. Vozmediano, and A. Cano,
Nat. Phys. {\bf 7}, 811 (2011).

\bibitem{mucha}
M. Mucha-Kruczynski, I.L. Aleiner, and V.I. Fal'ko,
Phys. Rev. B {\bf 84}, 041404 (2011).

\bibitem{ferrari}
A.C. Ferrari, J C. Meyer, V. Scardaci, C. Casiraghi, M. Lazzeri,
F. Mauri, S. Piscanec, D. Jiang, K.S. Novoselov, S. Roth, and
A.K. Geim,
Phys. Rev. Lett. {\bf 97}, 187401 (2006).

\bibitem{pisana}
S. Pisana, M. Lazzeri, C. Casiraghi, K.S. Novoselov, A.K. Geim,
A.C. Ferrari, and F. Mauri,
Nat. Mat. {\bf  6}, 198 (2007).

\bibitem{yan1}
J. Yan, Y. Zhang, P. Kim, and A. Pinczuk,
Phys. Rev. Lett. {\bf 98}, 166802 (2007).

\bibitem{yan2}
J. Yan, E.A. Henriksen, P. Kim, and A. Pinczuk,
Phys. Rev. Lett. {\bf 101}, 136804 (2008).

\bibitem{malard}
L.M. Malard, D.C. Elias, E.S. Alves, and M. A. Pimenta,
Phys. Rev. Lett. {\bf 101}, 257401 (2008).

\bibitem{kuzmenkoFano}
A.B. Kuzmenko, L. Benfatto, E. Cappelluti, I. Crassee,
D. van der Marel, P. Blake, K.S. Novoselov, and A.K. Geim,
Phys. Rev. Lett. {\bf 103}, 116804 (2009).

\bibitem{tang}
T.T. Tang, Y. Zhang, C.-H. Park, B. Geng, C. Girit, Z. Hao,
M.C. Martin, A. Zettl, M.F. Crommie, S.G. Louie, Y.R. Shen, and
F. Wang,
Nat. Nanotech. 5, {\bf 32} (2010).

\bibitem{lilui}
Z.Q. Li, C.H. Lui, E. Cappelluti, L. Benfatto, K.F. Mak, G.L. Carr,
J. Shan, and T.F. Heinz,
arXiv:1109.6367

\bibitem{review}
A.H. Castro Neto, F. Guinea, N.M.R. Peres, K.S. Novoselov,
and A.K. Geim,
Rev. Mod. Phys. {\bf 81}, 109 (2009).

\bibitem{zhang}
L.M. Zhang, Z.Q. Li, D.N. Basov, M.M. Fogler,
Z. Hao, and M. C. Martin,
Phys. Rev. B {\bf 78}, 235408 (2008).

\bibitem{koshino}
M. Koshino and E. McCann,
Phys. Rev. B {\bf 79},  125443 (2009).

\bibitem{kuzmenko2L}
A.B. Kuzmenko, I. Crassee, D. van der Marel, P. Blake, and
K.S. Novoselov,
Phys. Rev. B {\bf 80}, 165406 (2009).

\bibitem{li}
Z.Q. Li, E.A. Henriksen, Z. Jiang, Z. Hao, M.C. Martin,
P. Kim, H.L. Stormer, and D.N. Basov,
Phys. Rev. Lett. {\bf 102}, 037403 (2009).

\bibitem{avetisyan}
A.A. Avetisyan, B. Partoens, and F.M. Peeters,
Phys. Rev. B {\bf 81},  115432 (2010).

\bibitem{mcd}
F. Zhang, B. Sahu, H. Min, and A.H. MacDonald,
Phys. Rev. B {\bf 82},  035409 (2010).

\bibitem{zou}
K. Zou, X. Hong, and J. Zhu,
Phys. Rev. B {\bf 84}, 085408 (2011).

\bibitem{lui}
C.H. Lui, Z.Q. Li, K.F. Mak, E. Cappelluti, and T.F. Heinz,
arXiv:1105.4658v1.

\bibitem{apalkov}
V.M. Apalkov and T. Chakraborty,
arXiv:1111.3580v1.

\bibitem{dresselhaus}
M.S. Dresselhaus and G. Dresselhaus,
Adv. Phys. {\bf 51}, 1 (2002).

\bibitem{ando1}
T. Ando,
J. Soc. Phys. Jpn. {\bf 75}, 124701 (2006).

\bibitem{ando2}
T. Ando,
J. Soc. Phys. Jpn. {\bf 76}, 104711 (2007).

\bibitem{ando3}
T. Ando and M. Koshino,
J. Soc. Phys. Jpn. {\bf 78}, 034709 (2009).

\bibitem{gava}
P. Gava, M. Lazzeri, A.M. Saitta, and F. Mauri,
Phys. Rev. B {\bf 80}, 155422 (2009).

\bibitem{our}
E. Cappelluti, L. Benfatto, and A.B. Kuzmenko,
Phys. Rev. B {\bf 82}, 041402 (2010).

\bibitem{notedirac}
We mean here with ``Dirac point'' any point where
upper and lower bands touch each other, in a semimetal way,
independently of their linear or parabolic (or higher order) nature.


\bibitem{manes}
J.L. Ma\~{n}es, F. Guinea, and M.A. H. Vozmediano,
Phys. Rev. B {\bf 75}, 155424 (2007).

\bibitem{samsonidze}
Ge. G. Samsonidze, E.B. Barros, R. Saito, J. Jiang,
G. Dresselhaus, and M. S. Dresselhaus,
Phys. Rev. B {\bf 75}, 155420 (2007).

\bibitem{khan}
F.S. Khan and P.B. Allen,
Phys. Rev. B {\bf 29}, 3341 (1984)

\bibitem{lazzeri}
M. Lazzeri, C. Attaccalite, L. Wirtz, and F. Mauri,
Phys. Rev. B {\bf 78}, 081406 (2008).

\bibitem{piscanec}
S. Piscanec, M. Lazzeri, F. Mauri, A.C. Ferrari, and J. Robertson,
Phys. Rev. Lett. {\bf 93}, 185503 (2004).

\bibitem{mariani}
E. Mariani, A.J. Pearce, and F. von Oppen,
arXiv:1110.2769.

\bibitem{manzardo}
M. Manzardo, E. Cappelluti, and A.B. Kuzmenko,
unpublished.

\bibitem{espresso} 
P. Giannozzi {\it et al.},
J. Phys. Condens. Matter {\bf 21}, 395502 (2009).

\bibitem{pz} 
Appendix C of J.P. Perdew and A. Zunger, 
Phys. Rev. B {\bf 23}, 5048 (1981).

\bibitem{pp}
B. Partoens and F.M. Peeters,
Phys. Rev. B {\bf 74}, 075404 (2006).

\bibitem{gruneis}
A. Gr\"uneis, C. Attaccalite, L. Wirtz, H. Shiozawa,
R. Saito, T. Pichler, and A. Rubio,
Phys. Rev. B {\bf 78}, 205425 (2008).

\bibitem{mccann}
E. McCann and V.I. Fal'ko,
Phys. Rev. Lett. {\bf 96}, 086805 (2006).


\end{thebibliography}
\end{document}